\documentclass[manuscript]{aastex}
\usepackage{graphicx}
\begin{document}
\title{Long GRBs and massive stellar explosions from frame dragging around rotating black holes} 
\author{Maurice H.P.M. van Putten}
\email{mvp@sejong.re.kr}
\affil{Department of Astronomy, Sejong University, 98 Gunja-Dong Gwangin-gu, Seoul 143-747, Korea}

\begin{abstract}
The most energetic GRB-supernovae probably derive from rotating stellar mass black holes. Based on BeppoSax data, we identify a mechanism for exploding a remnant stellar envelope by disk winds. A specific signature is high frequency modulations in the accompanying prompt GRB emission from dissipation in high energy emissions along the black hole spin axis due, in part, to forced turbulence in the inner disk or torus mediated by frame dragging. A majority of long GRBs are found to have significant autocorrelation below 10 Hz with chirps extending up to 1000 Hz. Their comoving Fourier spectra satisfy a power law with index $\alpha\simeq-0.82$ up to about one hundred Hz and comoving chirp spectra show broken power laws with $\alpha\simeq-0.65$ up to 10 Hz, $\alpha\simeq-0.25$ up to a few hundred Hz and $\alpha\simeq0$ beyond. These high frequency signatures are the most direct signature of long-lived turbulence down to the ISCO of rotating black holes, pointing directly accompanying long duration bursts in gravitational wave emission.
\end{abstract}


\maketitle

\section{Introduction}

Long GRBs and core-collapse supernovae are the most energetic transients in the Universe associated with stellar mass black holes 
and (proto-)neutron stars. The energy in long GRBs may reach one solar mass-energy in isotropic equivalent emission, allowing them 
be seen close to the era of reionization \citep{sal09}. The beamed ultra-relativistic outflows creating the observed GRB-afterglow emissions 
and the generally aspherical stellar explosions of relatively massive stars \citep{mae08} point to an explosion process powered by an angular momentum 
rich inner engine \citep{bis70}. CC-SNe show broad and distinct distributions in narrow- and broad-line events \citep{mau10} that may 
represent different energies, possibly associated to black holes or neutron stars. Furthermore, a number of long GRBs in the {\em Swift} 
catalogue appear with no association to massive stars or supernovae (e.g. \cite{van11} for a compilation) with diverse X-ray afterglow 
emissions \citep{ber12,mar13}. 

Even though the fraction of core-collapse supernovae producing long GRBs is quite small \citep{fra01}, hyper-energetic GRB-supernovae such 
as GRB031203/SN2003lw and GRB03029/SN2003dh stand out in revealing an energy reservoir which far exceeds the maximal spin 
energy $E_* \simeq 3\times 10^{52}$ erg of rotating (proto-)neutron stars \citep{van11}. 
These events are likely to be of a more exotic origin, representing a massive release of energy of rapidly rotating black holes exceeding 
$E_*$ by up to two orders of magnitude. Rotating black holes naturally form in- and outside star-forming regions, from stellar progenitors in 
intra-day stellar binaries \citep{pac98} as well as mergers of neutron stars with another neutron stars \citep{bai08}. A similar outcome is 
expected from the tidal break-up of a neutron star around a black hole companion \citep{pac91}, where the latter may have a diversity in 
spin depending on the formation history \citep{van99}.

The durations of long GRBs are consistent with the lifetime of spin of black holes interacting with high density matter via an inner torus 
magnetosphere\citep{van99,van01a}. This picture unifies long GRBs from CC-SNe and mergers with rapidly spinning black holes, i.e., in mergers 
of neutron stars with black holes or another neutron star \citep{van08a,cai09}. In contrast, the durations of short GRBs are consistent with the 
time scale of hyper-accretion onto slowly spinning black holes with possibly (weak) X-ray afterglows \citep{van01a}. This prediction was confirmed 
by the detection of faint X-ray afterglows to the short Swift event GRB 050509B and the short HETE II event GRB 050709. A fair number 
of X-ray afterglows to short GRBs have since been identified that support a connection with long GRBs \citep{mar13}.

The physical process enabling rotating black holes, newly formed in prompt core collapse in relatively massive stars, to power an ensuing 
aspherical explosion requires efficient conversion of spin energy into baryon-rich winds emanating from an inner disk close to the event horizon. 
This does not preclude a preceding explosion, perhaps ad interim forming a (proto-)neutron star.
Relativistic frame dragging is a universal agent predicted by general relativity, which enables causal interactions between the angular momenta in astronomical objects and their surrounding fields and matter. Around rotating, it may naturally drive multiwindow radiation processes.

The existence of frame dragging has recently been experimentally established in non-energetic interactions by the two complementary satellite 
experiments LAGEOS II and Gravity Probe B \citep{ciu04,eve11}. Scaling of the GP-B measured value of $\omega$=-39 mas/year according to 
$\omega\simeq 2J/r^3$ shows a corresponding measurement at a distance $r\simeq 5$ million gravitational radii around an extremal black 
hole with the same angular momentum $J$ as the Earth. The present measurements present the non-relativistic limit of frame dragging in the
immediate vicinity of black holes described by (\ref{EQN_OM1}). According to the Kerr metric \citep{ker63}, the frame dragging angular 
velocity $\omega$ near a black hole of mass $M$ and angular 
velocity $\Omega_H$ scales effectively as
\begin{eqnarray}
\frac{\omega}{\Omega_H}\simeq \frac{1}{[1+\frac{r-r_H}{2M}]^3}.
\label{EQN_OM1}
\end{eqnarray} 
Here, $\Omega_H=(1/2M)\tan(\lambda/2)$ is the limit of $\omega$ at the horizon radius $r_H = 2M\cos^2(\lambda/2)$ where $\sin\lambda = J/M^2$
for a black hole with angular momentum $J$. 

We consider the exposure of a rotating black hole to an inner torus magnetosphere with variance $\sigma_B^2$ in poloidal magnetic field, resulting from aforementioned catastrophic events. The black hole develops a lowest energy state in equilibrium with an inner torus magnetosphere, described by an equilibrium magnetic moment which preserves a maximal horizon flux at all spin rates. Consequently, frame dragging 
mediates a dominant output of spin energy with luminosity \citep{van99}
\begin{eqnarray}
L_H \propto \sigma_B^2M^2\left(\Omega_H-\Omega_T\right)\Omega_T
\label{EQN_EM1}
\end{eqnarray}
by surrounding high density matter with angular velocity $\Omega_T$. $L_H$ is catalytically converted into various emission channels. A substantial
fraction will appear in powerful magnetic winds, that may drive expulsion of a remnant stellar envelope following black hole formation in core-collapse
of a massive star. While the black hole spins rapidly, the inner disk hereby morphs into a torus in suspended accretion with forced turbulent motions due to the competing torques acting on its inner and outer face \citep{van99,van01a}. In mergers, the same magnetic winds from would produce a powerful 
extragalactic radio burst.

In its lowest energy state, the equilibrium horizon flux can support an open magnetic flux tube out to infinity, all the while in suspended accretion 
due to (\ref{EQN_EM1}). Along this tube, frame dragging drives angular momentum outflow with a potential energy \citep{van05}
\begin{eqnarray}
E=\omega J_p
\label{EQN_EM2}
\end{eqnarray}
on particles with angular momentum $J_p$. Applied to electron-positrons with angular momentum $J_p=eA_\phi$, where $e$ refers to their 
electric charge and $A_a$ denotes the electromagnetic vector potential in the space time around the black hole, (\ref{EQN_EM2}) assumes 
UHECR energies in superstrong magnetic fields, enabling the formation of baryon-poor ultra-relativistic jets (BPJ). 

Frame dragging induced high energy outflows and magnetic winds are inevitably correlated in strength and temporal properties. 
Since the observed high energy emission derives from dissipation downstream in the former (e.g. \cite{tav96}), GRB light curves are subject
to time variability in the equilibrium horizon flux in response fluctuations in the inner torus magnetosphere. The latter possibly subject to 
supperradiant instabilities \citep{van99} and at any rate subject to inherently time-variable dynamo by turbulent mass motions \citep{bal91}
in the inner disk or torus. For the poloidal magnetic field, the combined result is quantified by $\sigma_B^2$. These modulations are in addition to any other time-variability in long GRBs, as may arise from fluctuations in orientation of the collimating disk winds, as well as in-situ in the dissipative shock fronts downstream of the BPJ.

The temporal structure of GRB light curves hereby carry potentially significant information about the state of matter in the inner engine of 
GRB-supernovae. An exhaustive energy release from the black hole causes the black hole to spin down. By the corresponding decrease in
frame dragging according to (\ref{EQN_OM1}), it produces an underlying decay in the BPJ powering the prompt GRB emission. Such gradual 
decay can be seen in a normalized light curve of 1491 long GRBs of the BATSE catalogue \citep{van12a}. By the aforementioned temporal 
correlations between (\ref{EQN_EM1}) and ({\ref{EQN_EM2}), we anticipate, furthermore, signatures up to relatively high frequencies 
stemming from aforementioned turbulent mass motion in a torus about the inner most stable circular orbit (ISCO). These signatures will have
a broad band extension down to low frequency modulations of the collimating winds by mass motions in the inner disk further out.

Here, we set out to identify the anticipated broad band spectra in light curves of long GRBs by a novel analysis in the frequency domain. 
We analyze low and high frequency spectra of long GRBs from BeppoSax catalogue of 1082 GRBs from the Gamma-Ray Burst Monitor (GRBM, \cite{fro09}). GRBM recorded the first 8-10 seconds of each at a 2 kHz sampling rate in the energy range 40-700 keV in a Field of View (FOV) of about $2\pi$ sr \citep{fro09}. The GRBM offers a unique window to search for low and high frequency signatures up to 1 kHz covering (\ref{EQN_f}). We selected 72 bright GRBs with relatively large photon counts from the BeppoSax catalogue for analysis (Fig. \ref{fig:overview}). Our sample of 72 bursts has an approximate log-normal distribution in durations from 3 to 453 s. 
\begin{figure}[h]
\centerline{
\includegraphics[width=55mm,height=50mm]{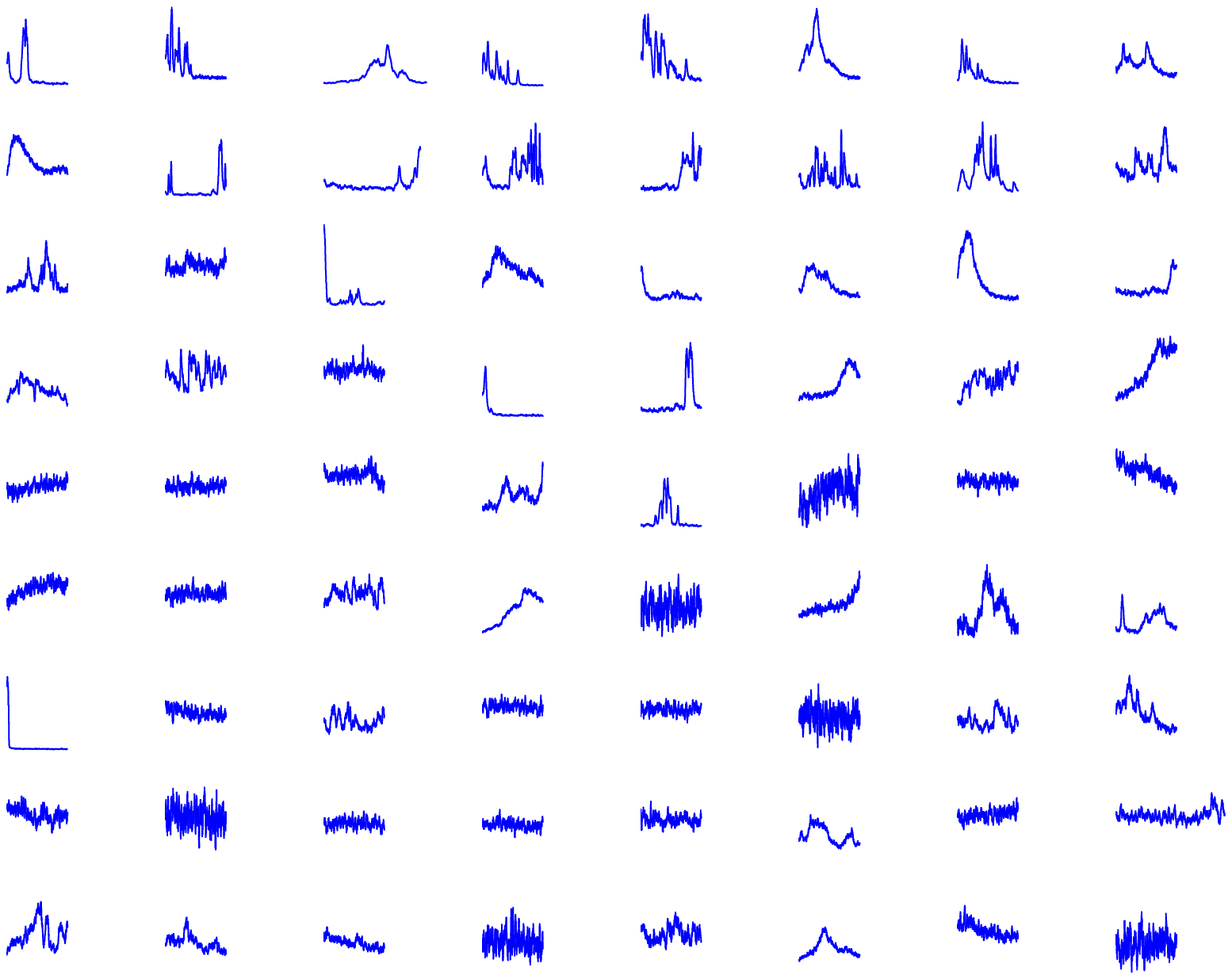}				
\includegraphics[width=55mm,height=50mm]{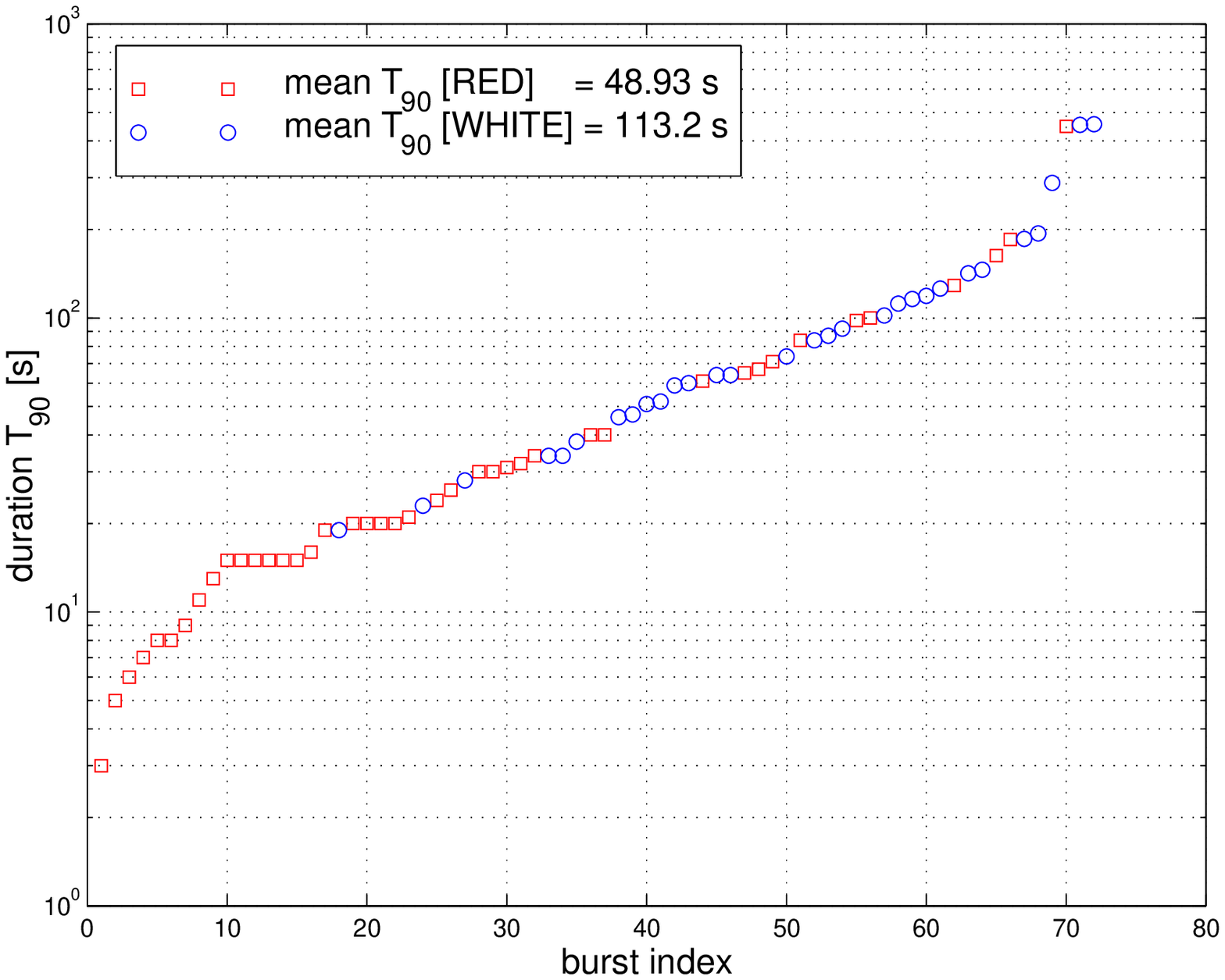}				
\includegraphics[width=55mm,height=50mm]{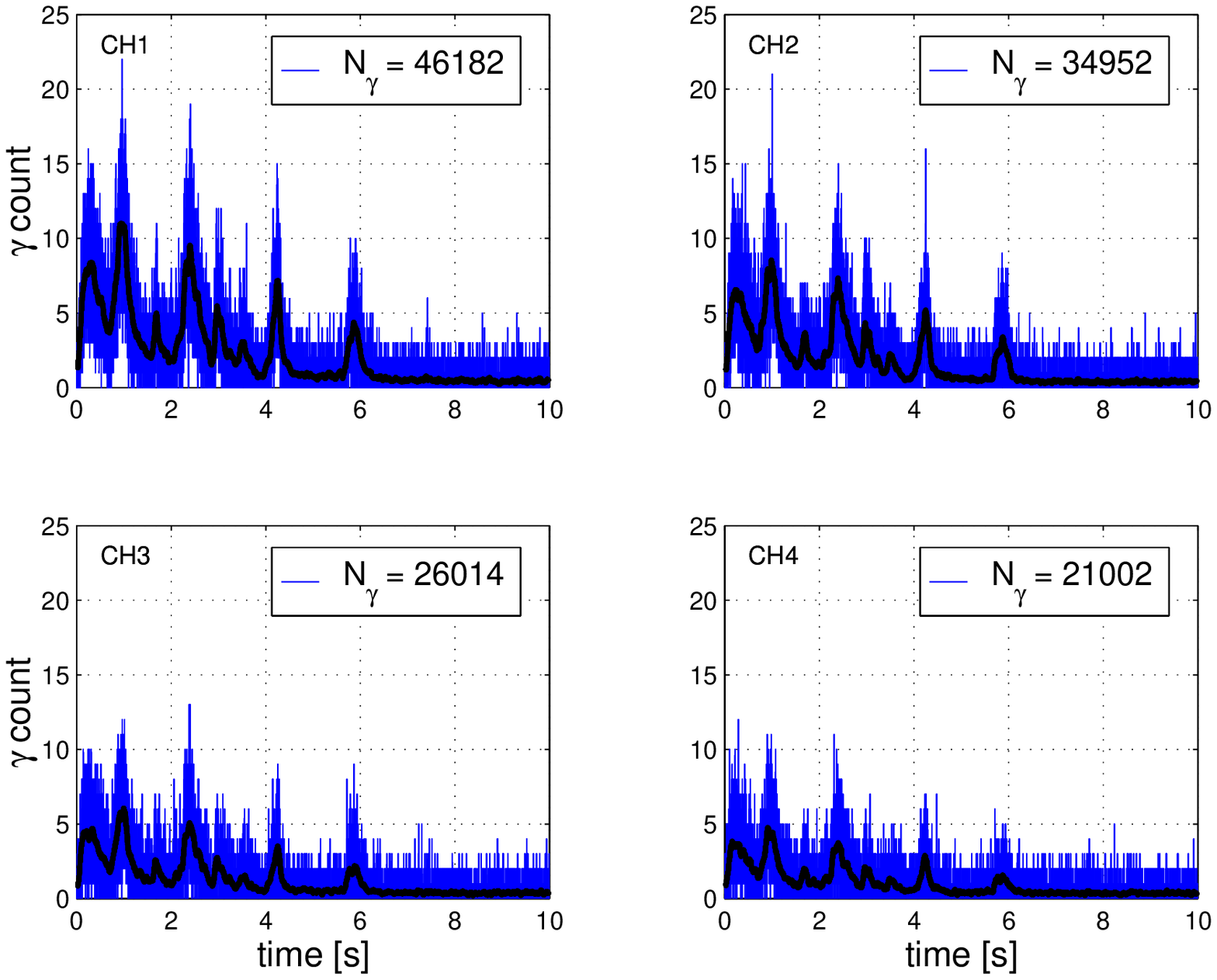}}				
\caption{Shown are the smoothed light curves ({\em left}, sorted by duration $T_{90}$) of an ensemble of 72 bright long GRBs in 
the BeppoSax catalogue from 3-456 s ({\em middle}, in two spectral colors, Fig. 2). Each GRB is sampled in four channels by identical BeppoSax detectors at 2 kHz over the first 10 s (8 s in a few events), here sorted by total photon counts as illustrated by GRB 010109 ($T_{90}=7$ s) in our ensemble ({\em right}, with superimposed smoothed light curve).}
\label{fig:overview}
\end{figure}

\begin{table}[h]
\caption{A BeppoSax sample of 72 bright long GRBs with durations $T_{90}=4-454$ s classified by the color (white/red) of their low frequency spectra.}
\center{
\begin{tabular}{llllllllllll}
\hline\hline
GRB 			& $T_{90}$ & color & GRB & $T_{90}$ & color & GRB & $T_{90}$ & color & GRB & $T_{90}$ & color\\
\hline\hline
970111   	& 31   	& R  & 971223 	&  47  &  W	& 990705 & 32 & R 		& 000630 & 26 & R\\
970116	& 112	& W &  980203 	&  23  &  W 	& 990718 & 126 &W	 	& 000718 & 34 & R\\
970117	& 13 		& R &  980306 &  21  &  R 	& 990913 & 40 & R		& 001004 & 9 & R\\
970315	& 15		& R &  980329 &  19   & W 	& 991116 & 185 & R 	& 001011 & 24 & R\\
970517	& 5		& R &  980428 & 100 & R 		& 991124 & 28 & W 		& 001212 & 67 & R\\
970601	& 30		& R &  980615 & 64   & W		& 991216 & 15 & R 		& 001213 & 454 & W\\
970612	& 38		& W &  980617 & 186 & W 	& 000115 & 15 & R 		& 010109 & 7 & R\\
970616	& 64		& W &  980706A & 71  & R 	& 000214 & 8 & R 		& 0101222 & 74 & W\\
970625	& 15		& R & 980706B & 146 & W 	& 000218 & 20 & R 		& 010317 & 30 & R\\
970627	& 15		& R &  980728 & 52 	 & W 		& 000226 & 84 & R 		& 010326 & 19 & R\\
970706	& 59		& W & 980827 & 51   & W 		& 000227 & 102 & W 	& 010408 & 4 & R \\
970816	& 6		& R & 981111 & 34  & W 		& 000323 & 46 & W 		& 010412 & 60 & W\\
971019	& 20		& R & 981203 & 142 & W		& 000327 & 87 & W 		& 010504 & 15 & R\\
971027	& 11 		& R & 990118 & 84  & W 		& 000328 & 116 & W 	& 010619 & 449 & R\\
971029	& 92		& W & 990123 & 61 & R		& 000419 & 20 & R 		& 010710 & 20 & R\\
971110  	& 194 	& W & 990128 & 8    & R 		& 000429 & 163 & R 	& 010826 & 288 & W\\
971114	& 98		& R & 990506 & 129 & R 		& 000528 & 65 & R		& 010922 & 40 & R\\
971208  	& 456	& W & 990620 & 16  & R 		& 000621 & 119 & W 	& 011003 & 34 & W\\
\hline\hline
\end{tabular}
\label{TABLE_1.1}}
\begin{tabular}{l}
\mbox{}\hskip0.01in
\end{tabular}
\end{table}

Fig. 2 shows a Fourier analysis of the ensemble of Fig. \ref{fig:overview} following the calculation of the autocorrelation coefficients (ACC) of the individual light curves. The ACC analysis reveals two-color spectra in the low frequency range, from essentially white to red up to about 10 Hz. In a lower frequency bandwidth, a similar partition has been found in the BATSE catalogue \citep{bor04}. We find correlations that extend to 62.5 Hz in a few bursts, defined by zero-crossing times following a high pass filter. The resulting 30 white and 42 red bursts show relatively featureless and, respectively, pronounced Fourier spectra with typical power law behavior in the infrared, i.e., up to about about 1 Hz and 10 Hz with index $\alpha \simeq 0.25$ and $\alpha\simeq-0.82$, respectively. These observations are consistent with the Kolmogorov spectrum of the average power density spectrum observed in the BATSE catalogue with a noticeable change in $\alpha$ for relatively weak bursts \citep{bel98}.

In our model, the observed duration $T_{90}$ of long GRBs is due to a diversity in redshift, $z$, in black hole masses, $M$, and a in $\sigma_B^2$ due to different masses $M_T$ of the torus. $T_{90}$ is proportional to the first two. While $T_{90}$ is inversely proportional to $\sigma_B^2$, set by the kinetic energy in the torus \citep{van03}. The ratio $M_T/M$ is probably a random variable independent of $z$ and $M$. The normalized frequency $f T_{90}/\bar{T}^\prime_{90}$, $T_{90}^\prime=T_{90}/(1+z)$, $\bar{T}_{90}=48.93$ s, is therefore a suitable proxy for a normalized comoving frequency. Since BeppoSax bursts are mostly devoid of redshift information, we shall use a mean redshift $z=3$ similar to that of {\em Swift} bursts \citep{jak05}. Fig. 2 shows the resulting estimated comoving spectrum. The observed power law is found to extends to tens of Hz, beyond that obtained by direct averaging. A similar extension is found in the redshift corrected average spectrum of {\em Swift} bursts \citep{gui12}. At higher frequencies, the spectrum is dominated by apparent white noise. In the frequency range considered, there is no significant evidence for quasi-periodic oscillations (QPOs), as may be expected from a strongly turbulent inner engine. 
\begin{figure}[h]												
\centerline{				
\includegraphics[width=40mm,height=40mm]{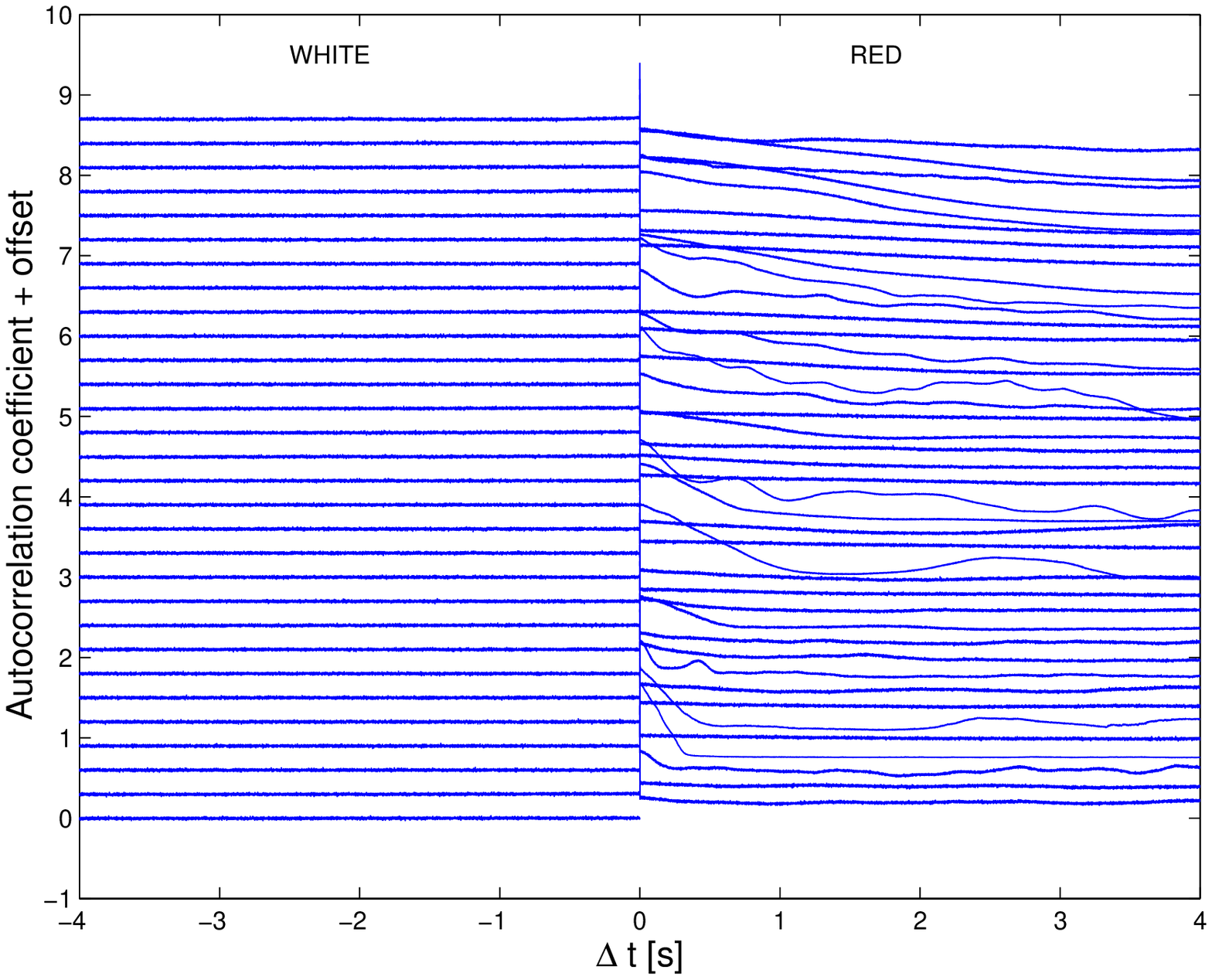}			
\includegraphics[width=40mm,height=40mm]{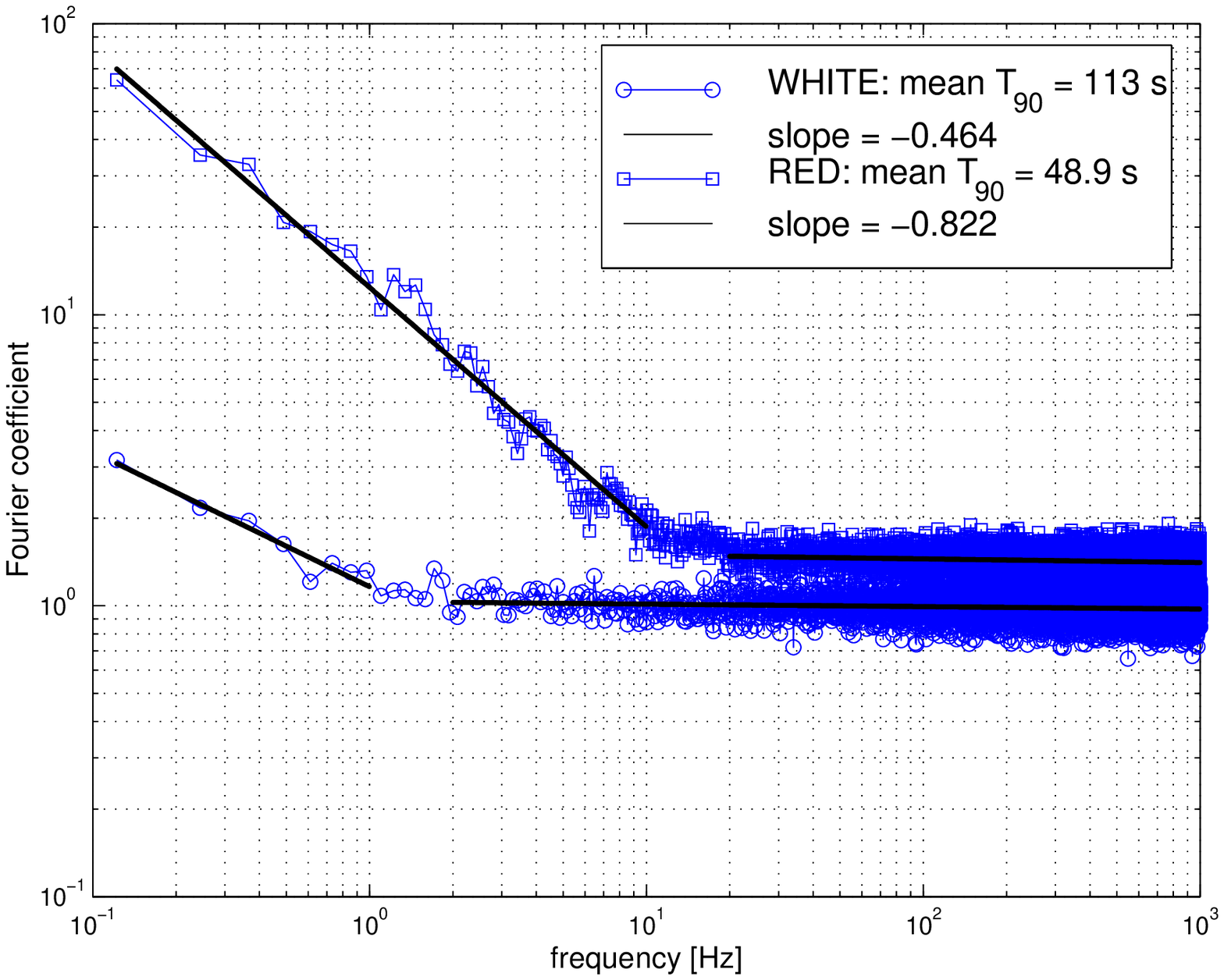}		
\includegraphics[width=40mm,height=40mm]{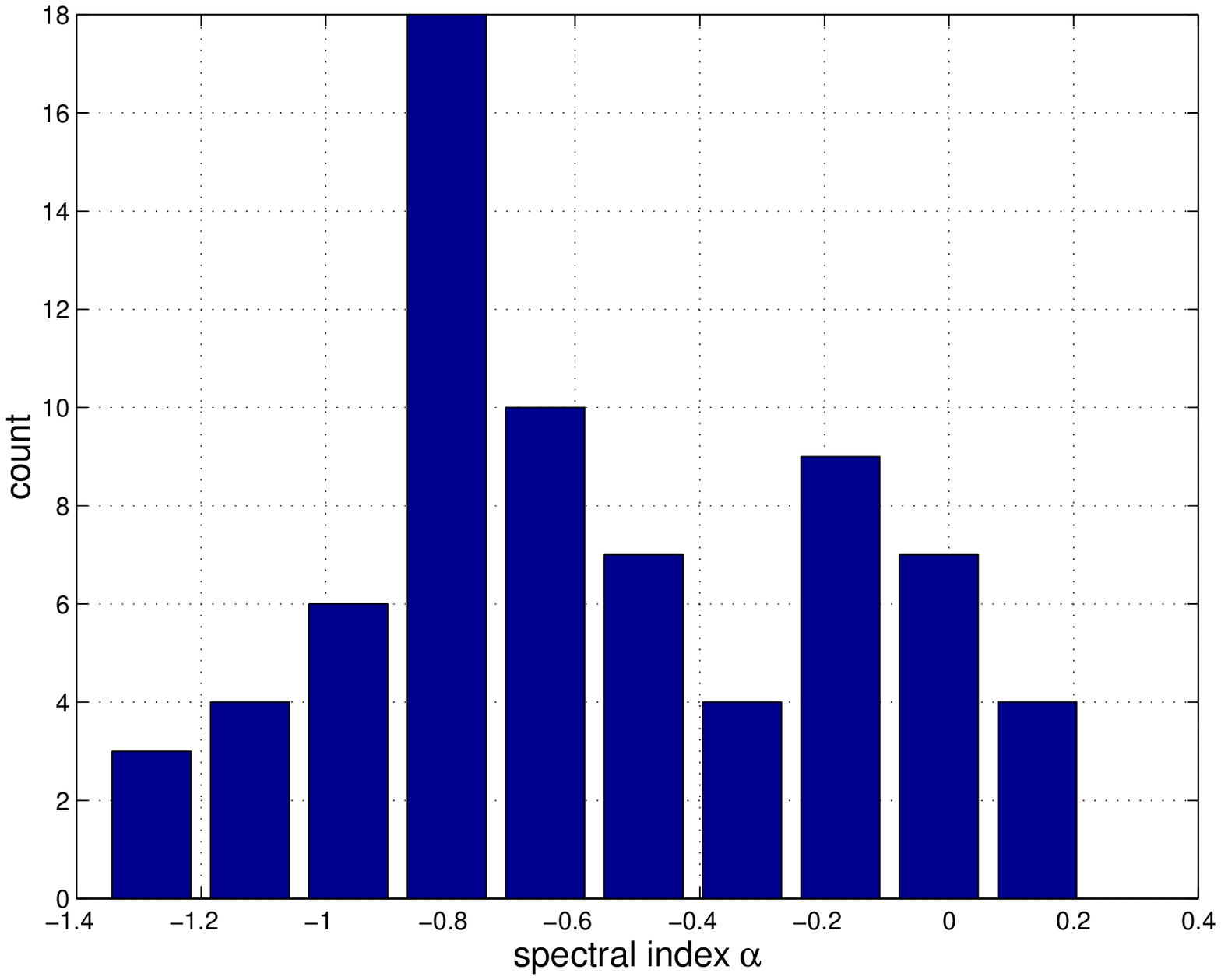}		
\includegraphics[width=40mm,height=40mm]{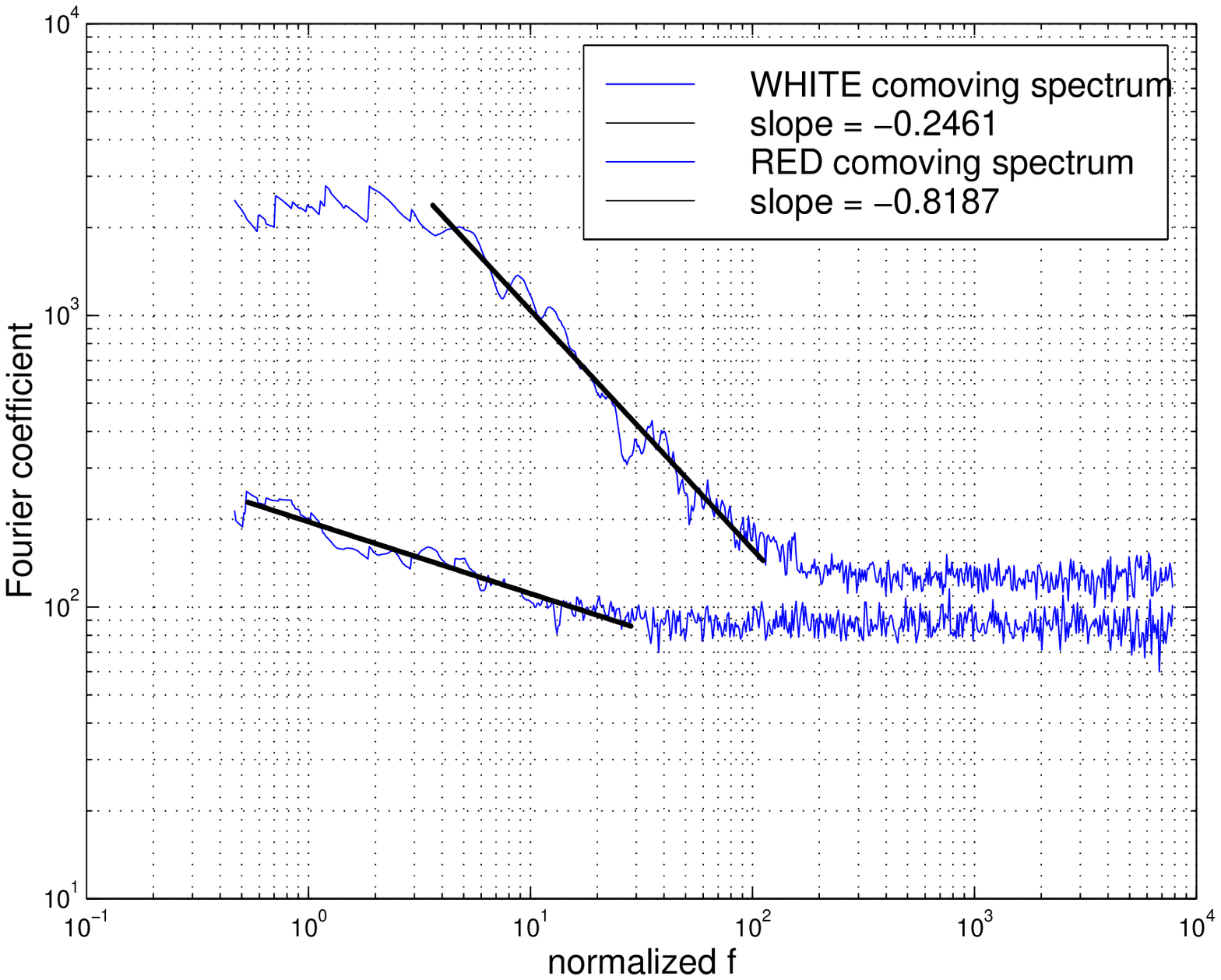}}		         		
\caption{({\em Left.}) The ensemble of 72 bursts contains relatively {white} and {red} bursts according to their autocorrelation coefficients and average Fourier frequency spectrum. ({\em Middle left.}) Their low frequency spectra show power law behavior $c(f)\propto f^{-\alpha}$ for their Fourier coefficients $c_i$ up to 1 Hz and, respectively, 10 Hz. They have mean photon counts of 0.5936 and 1.2569 per sample with mean $T_{90}$ of 113 s and 49 s, respectively, suggesting different distances. However, the indices $\alpha=-0.464$ and $-0.822$ ({\em straight lines}) show that these bursts are intrinsically different as part of an approximately bimodal distribution (over 0-5 Hz). The spectra shown are normalized such that the high frequency spectrum of the white bursts has mean 1, where it appears white Gaussian with  means 1, 1.4352 and $\sigma=0.0960,0.1221$, respectively. ({\em Right.}) The estimated comoving spectra of white
and red bursts show an extension of power law behavior up to about ten and, respectively, about 100 Hz with $\alpha=-0.2461$ and $-0.8187$.} 
\label{fig:color}
\end{figure}

To study high frequency spectra, we developed a novel broad-band chirp search by matched filtering, focused on signals with slowly varying frequencies with zero sensitivity to steady state frequencies.  Chirps are hereby different from quasi-periodic oscillations (QPOs), that pertain to frequencies the fluctuate around a steady mean. In our search, the highest frequency chirps are set by matter at the ISCO. As the black hole spins down, the ISCO expands in relaxation towards a Schwarzschild black hole \citep{van08a} causing this frequency limit to gradually decreases in time. For an initially maximally spinning Kerr black hole with initial mass $M$, the observed frequency of such intermittent high frequency chirps about the ISCO produced by a source at redshift $z$ satisfies \citep{van11a}
\begin{eqnarray}
f_{chirp}(t)=  \left[0.1747+0.5743 e^{-7.5\,t/T}\right]\left(\frac{4}{1+z}\right)\left(\frac{m}{2}\right) \left(\frac{10M_\odot}{M}\right)\mbox{kHz}
\label{EQN_f}
\end{eqnarray}
for a multipole mass-moment $m$, where $T$ denotes the burst duration $T_{90}$. The actual late-time frequency has a range of about 20\% the value of which depends on the initial spin. The late time comoving frequency for $m=1$ is about 300-350 Hz.

Following \citep{van11a}, our templates are obtained as {\em slices} of intermediate duration $\tau = 1$ s of a long duration chirp with exponential decay of the form (\ref{EQN_f}). Since we are interested in a model-independent search for chirps with possibly randomly decreasing or increasing frequencies, we form model templates of differences of sliced templates forwards and backwards in time. We next calculate sample cross-correlation coefficients (SCC) between these difference templates and data over a broad range of model parameters, here frequency and rate of change of frequency. Since the SCC comprises relatively large number of samples (2048 for $\tau=1$ s), it generally produces a near-Gaussian distribution with unit variance by the central limit theorem. The significance of a choice of model parameters is therefore given by the maximum of the absolute value of its SCC, here referred to as the signal-to-noise ratio (SNR). In light of (\ref{EQN_f}), we consider a chirp search with uniform distributions of the logarithm of frequency, covering 4-1000 Hz, and the time scale $T$ for frequency change, covering 6-600 s. The analysis is carried out over the first 8 seconds ($2^{14}$ samples) of BeppoSax data, common to all bursts in our ensemble. 

In light of the considerable discretization noise due to the extremely low photon counts on the order of 1 photon per sample on average in the 2 kHz BeppoSax light curves, we employed redundant controls, by calculating our results relative to those obtained from the same light curves after randomization in time and against light curves generated by a Gaussian random number generator. Here, time randomization serves to destroy any autocorrelation, while leaving invariant total power and histograms of photon counts. By way of example, Fig. \ref{fig:delta1} shows the results for GRB010109 and the resulting chirp spectrum, expressing SNR in excess of that obtained in our control following time randomization. For ensemble averages, these two controls are effectively the same, although some differences may appear in the analysis of individual light curves.
\begin{figure}[h]
\centerline{
\includegraphics[width=65mm,height=55mm]{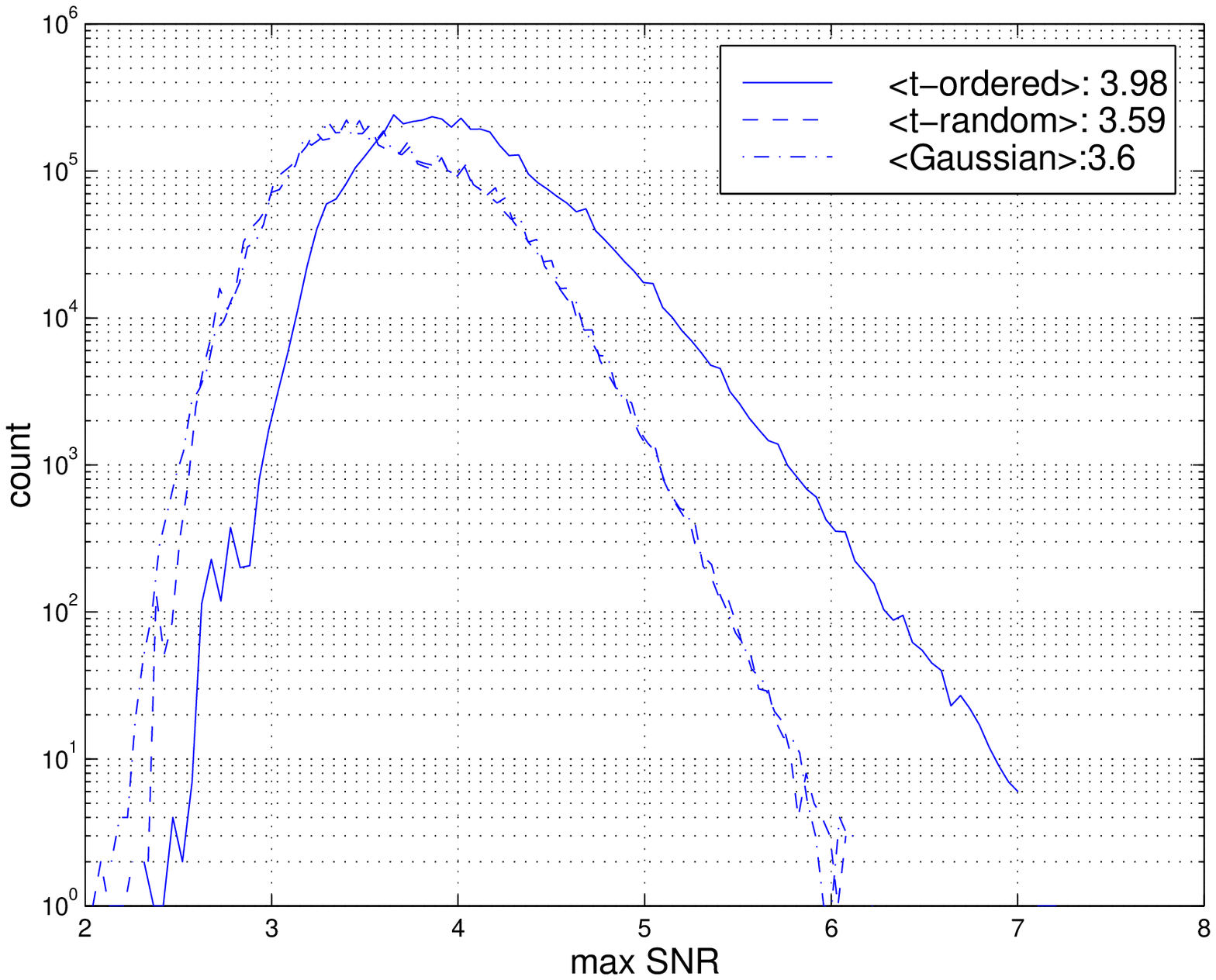} 	
\includegraphics[width=65mm,height=55mm]{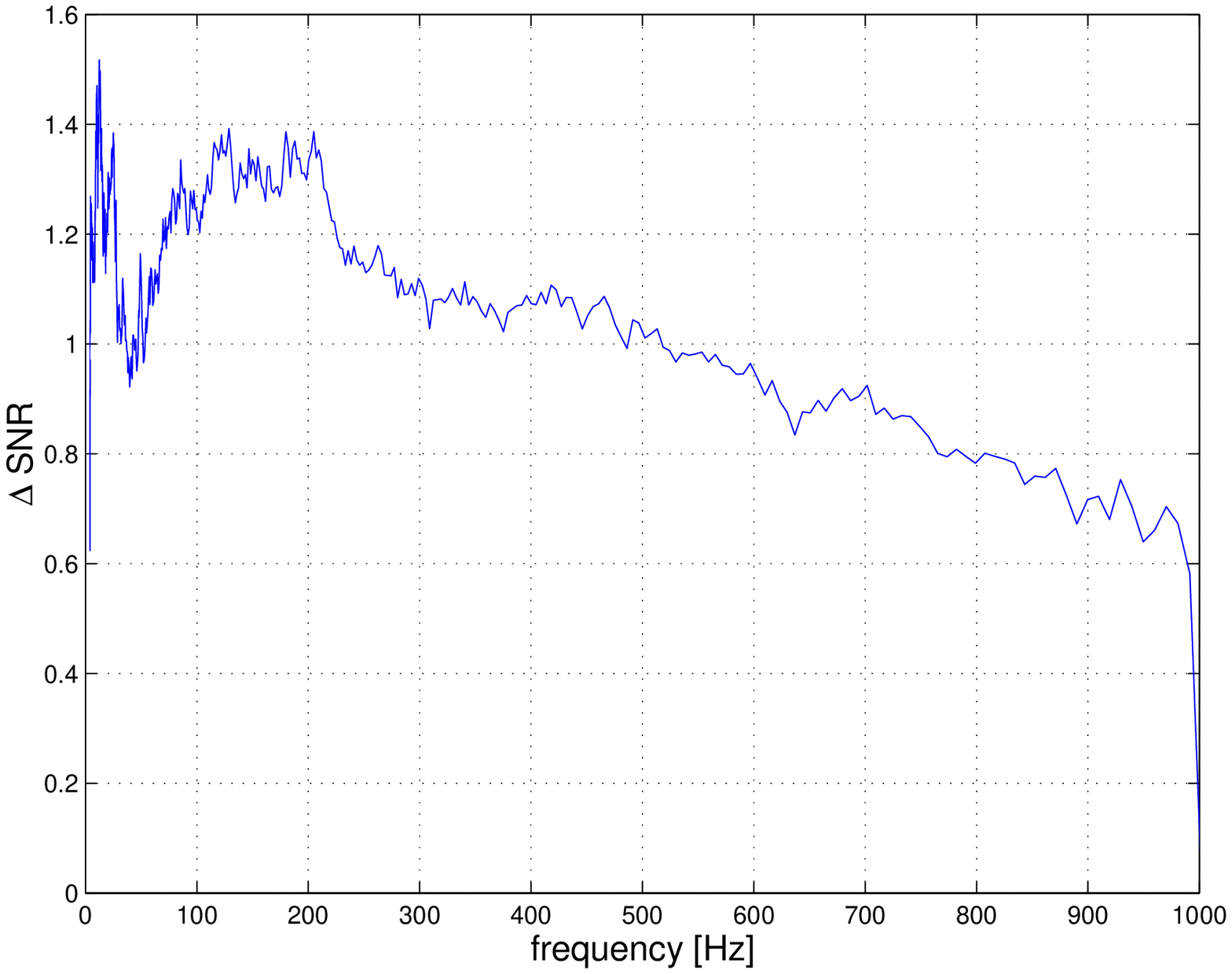} 	
}
\caption{
({\em Left.}) The distributions of max SNR of each chirp template are shown
for a range of model parameters for GRB010109, obtained from matched filtering applied 
to the original (time-ordered) light curve, the time-randomized light curves and a light curves 
generated by a Gaussian random number generator. 
The distribution of the max SNR for time-ordered light curves is shifted relative to 
those of the latter two. A chirp spectrum obtains from the difference $\Delta$ SNR in their mean differentiated
by template frequency, providing a measure for the strength of temporal modulations in the light curve ({\em right.}).}
\label{fig:delta1}
\end{figure}

Similar to the spectral index $\alpha$ in Fig. \ref{fig:color}, we express the color of the low frequency 
spectra by the logarithmic ratio of Fourier coefficients $c_i$ over the first 8 s ($N=2^{14}$ samples),
\begin{eqnarray}
\mbox{color~}=\frac{\sum_{i=1}^{P} \log |c_i|}{\sum_{i=N-Q}^{N}\log |c_i|},
\end{eqnarray}
where the numerator sum over $P=40$ covers the frequency range 0-5 Hz, normalized to the mean of the asymptotically flat spectrum 
using $Q=1000$. The color thus expressed in the low frequency range points to the spectrum in the high frequency range, as anticipated
from the maximum in $\Delta$ SNR across all templates, as shown in Fig. \ref{fig:delta2}.

Fig. \ref{fig:delta2} shows a continuation of the white and red groups identified in the Fourier analysis shown in Fig. \ref{fig:color} into the high 
frequency chirp spectrum up to 1000 Hz. The results establish that the red-white dichotomy extends throughout the low and high frequencies 
up to 1000 Hz.
\begin{figure}[h]
\centerline{
\includegraphics[width=54mm,height=50mm]{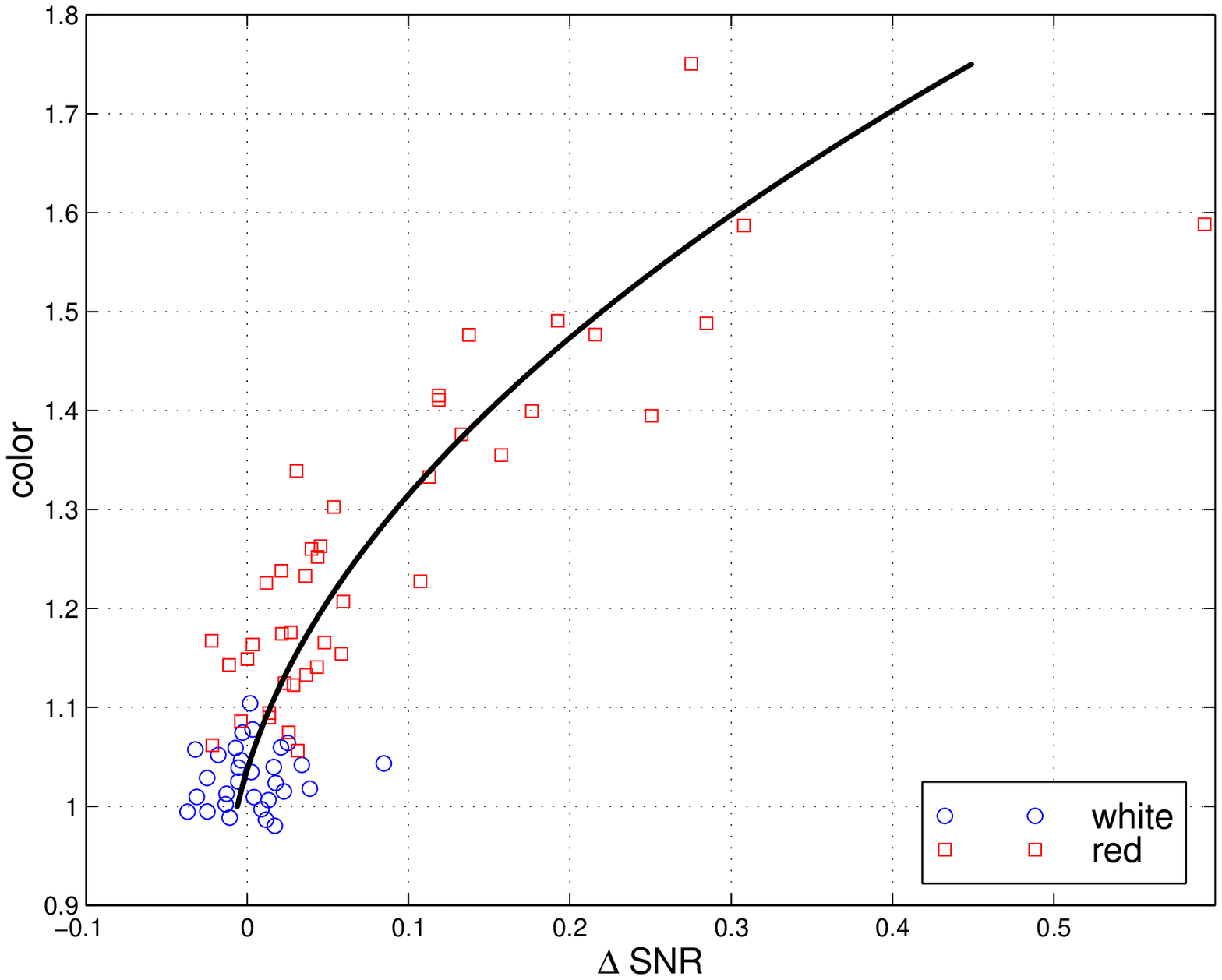}		
\includegraphics[width=54mm,height=50mm]{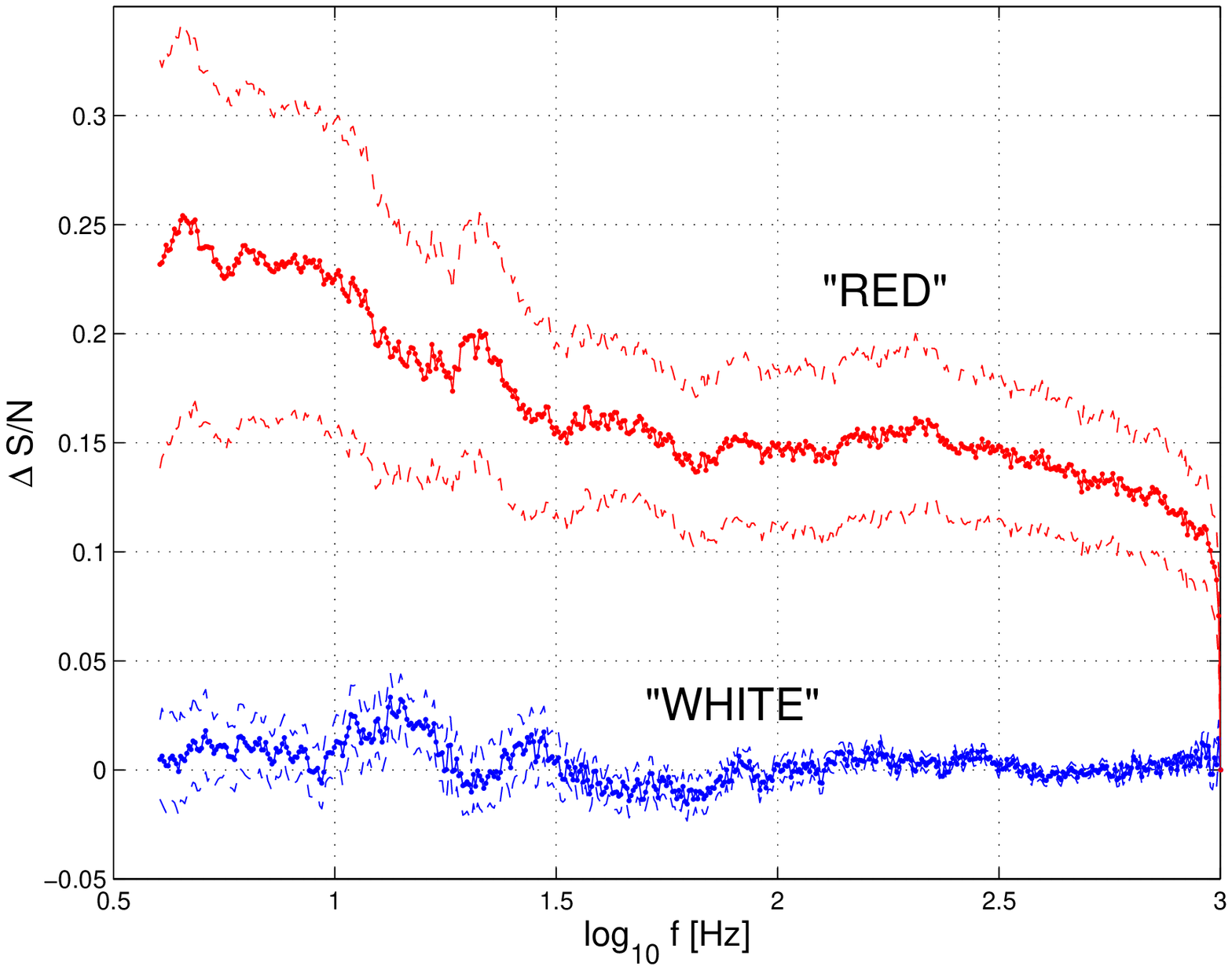} 			         
\includegraphics[width=54mm,height=50mm]{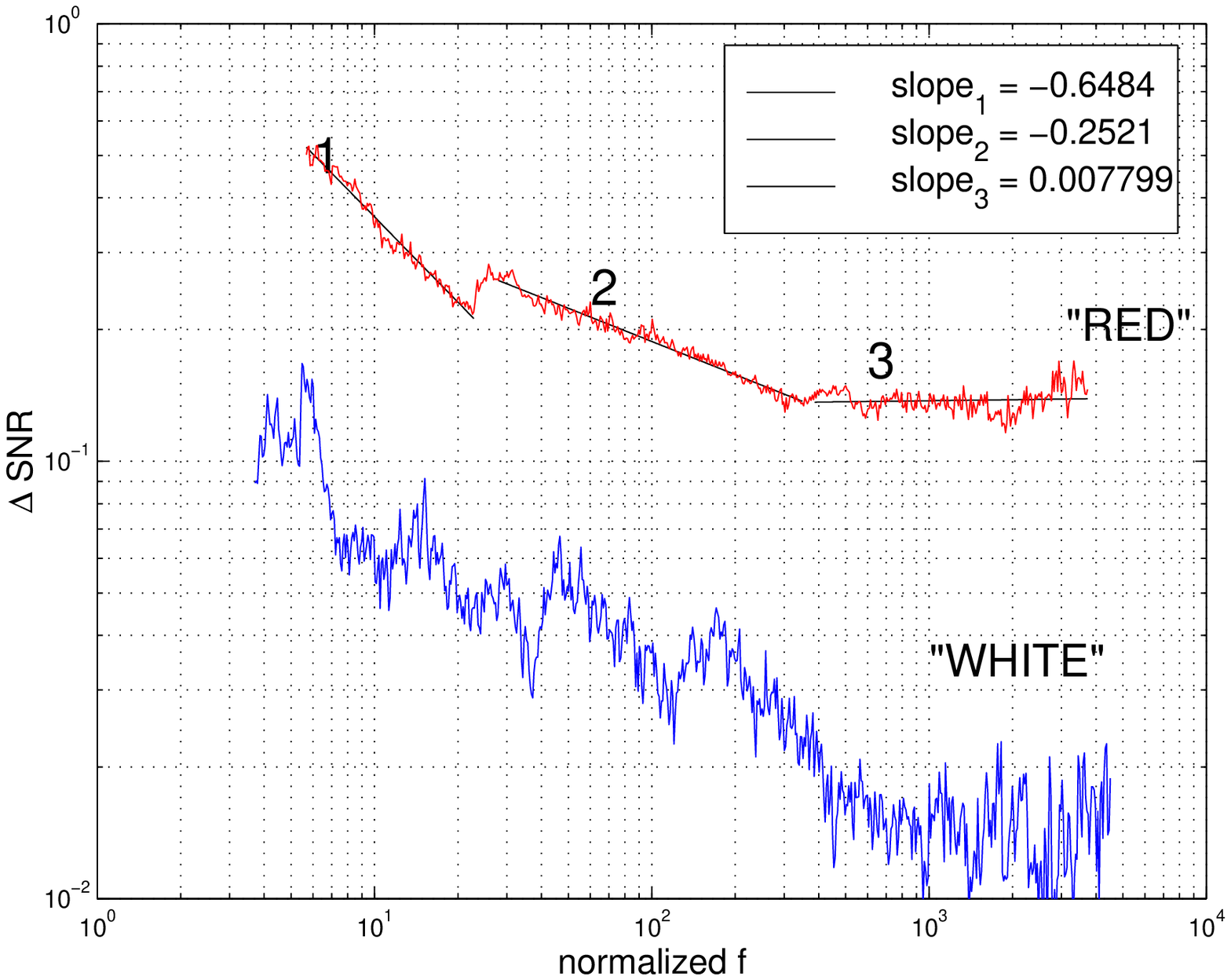}}		         
\caption{
({\em Left.}) High frequency chirps are correlated with low frequency 
modulations through a correlation of color with the maximum $\Delta$ SNR in each GRB as seen in the brightest detector channel.  
({\em Middle.}) 
The chirp spectrum over 0-1000 Hz is shown as an ensemble average for the red and white bursts. The dashed lines indicate the standard error in the mean. While the ensemble spectrum of white busts is featureless, that of red decays over a few hundred Hz followed by a smooth break to linear decay up to about 900 Hz. These results are calculated using matched filtering over 5.12 million templates. ({\em Right.}) The normalized spectrum shows a typical power law behavior, here with $\alpha\simeq - 0.65$ to $f_1\simeq 10-20$ Hz, $\alpha\simeq -0.25$ up to $f_2\simeq 300$ Hz and $\alpha\simeq 0$ beyond. The associated spectrum of the chirp time scale ratio $T/T_{90}$ is essentially uniform (not shown).}
\label{fig:delta2}
\end{figure}

Our frequency analysis reveals a group of red bursts with power law behavior in their Fourier spectra and normalized chirp spectra, where the
latter extends up to frequencies well beyond 1000 Hz in the comoving frame of the bursts. The break at $f_2\simeq 300$ Hz in the 
power law behavior of the normalized comoving chirp spectrum is consistent with $m=1$ at the ISCO in the de-redshifted frequencies of (\ref{EQN_f}).

For red bursts, we attribute these extraordinary high frequencies to turbulent mass motions in the inner-most region of the inner engine, close to the 
ISCO around rotating black hole forced by frame dragging. These modulations are observed in the prompt GRB emission by 
correlations between (\ref{EQN_EM2}) and (\ref{EQN_EM2}) due to $\sigma_B^2$. Lower frequency modulations in the spectrum 
may derive from mass motions further out, by their modulation of disk winds collimating the BPJ about
 the axis of rotation of the central black hole. 

Sustained turbulence in high density matter around a black hole requires a tremendously powerful, continuous energy input, 
pointing to an enormous energy reservoir - here most naturally the extraordinary large energy reservoir in angular momentum
of a rapidly rotating Kerr black hole. Our result should be contrasted with a standard thin disk in Keplerian rotation, which would 
be quiescent (non-turbulent).  

In the process of exhausting the spin energy of a rotating black hole, it slows down in relaxation to a slowly spinning black hole
described by the Schwarzschild metric. Spindown is hereby accompanied by an expansion of the ISCO. Turbulent motions and 
possibly accompanying non-axisymmetric instabilities hereby inevitably produce {\em negative chirps} in gravitational waves, 
that should be detectable in some of the hyper-energetic CC-SNe up to distances of about $D=35$ Mpc in the Local Universe 
by the upcoming advanced detectors LIGO-Virgo and KAGRA \citep{van11a}. A detection allows for calorimetric identification of 
Kerr black holes as the most exotic inner engines in nature \citep{van02}.

{\bf Acknowledgment.} The BeppoSax mission was an effort of the Italian Space Agency ASI with participation of The Netherlands Space Agency NIVR. Some of the calculations were performed at the CAC/KIAS, KISTI, XSEDE/NSF and VPGEONET. The author thanks F. Frontera and C. Guidorzi for kindly providing a sample of high resolution BeppoSax data and M. Della Valle for valuable comments on the manuscript.


\begin{thebibliography}{99}
\bibitem[Salvaterra et al.(2011)]{sal09} Salvaterra, R., Della Valle, M., Campana, S., et al., 2009, Nature, 461, 7268;
Tanvir, N.R., Fox, D.B., Levan, A.J., et al., 2009, Nature, 461, 7268; Ciucchiara, A., Levan, A.J., Fox, D.B., et al., 2011, ApJ, 736, 7
\bibitem[Maeda et al.(2008)]{mae08} Maeda, K., Kawabata, K., Mazzali, P.A., et al., 2008, Science, 319, 5667; 
Taubenberger, S., Valenti, S., Benetti, S., et al., 2009, MNRAS, 397, 677
\bibitem[Bisnovatyi-Kogan(1970)]{bis70} Bisnovatyi-Kogan, G. S. 1970, Astron. Zh., 47, 813
\bibitem[van Putten et al.(2011a)]{van11} van Putten, M.H.P.M., Kanda, N., Tagoshi, H., Tatsumi, D., Masa-Katsu, F., \& Della Valle, M., 2011, Phys. Rev. D, 83, 044046
\bibitem[Maurer et al.(2010)]{mau10} Maurer, J.I., Mazzali, P.A., Deng, J., et al., 2010, MNRAS, 402, 161 
\bibitem[Bernardini et al.(2012)]{ber12} Bernardini, M.G., Margutti, R., Zaninoni, E., \& Chincarini, G., 2012, MNRAS, 425, 1199
\bibitem[Margutti et al.(2013)]{mar13} Margutti, R., Zaninoni, E., Bernardini, M.G., et al., 2013, MNRAS, 428, 729
\bibitem[Frail et al.(2001))]{fra01} Frail, D.A., et al., 2001, ApJ, 562, L55; van Putten, M.H.P.M., \& Regimbau, T., 2003, ApJ, 593, L15; 
Guetta, D., Piran, T., \& Waxman, E., 2005, ApJ, 619, 412; Guetta, D., \& Della Valle, M., 2007, ApJ, 657, L73
\bibitem[van Putten et al.(2011b)]{van11a} van Putten, M.H.P.M., Della Valle, M., \& Levinson, A., 2011, A\&A, 535, L6
\bibitem[Baiotti et al.(2008)]{bai08} Baiotti, L., Giacomazzo, B., \& Rezzolla, L., 2008, Phys. Rev. D, 2008, Phys. Rev. D, 78, 084033
\bibitem[Paczynski(1991)]{pac91} Paczynski, B.P., Acta Astron. 41, 257 (1991).
\bibitem[Paczy\'ski(1998)]{pac98} Paczy\'nski, B.P., 1998, ApJ, 494, L45
\bibitem[van Putten \& Gupta(2009)]{van09a} van Putten, M.H.P.M., \& Gupta, A.C., 2009, MNRAS, 394, 2238
\bibitem[van Putten(2012)]{van12a} van Putten, M.H.P.M., 2012, Prog. Theor. Phys., 127, 331
\bibitem[van Putten(1999)]{van99} van Putten, M.H.P.M., 1999, Science, 294, 115; ibid. 2009, MNRAS,  396, L81
\bibitem[van Putten(2005)]{van05}van Putten, M.H.P.M., 2005, Nuov. Cim., 28, 597; ibid. 2008, ApJ, 685, L63
\bibitem[Balbus \& Hawley(1991)]{bal91} Balbus, S.A., \& Hawley, J.F., 1991, ApJ, 376, 214; Hawley, J.F., \& Balbus, S.A., 1991, ApJ, 376, 223
\bibitem[van Putten \& Levinson(2003)]{van03} van Putten, M.H.P.M., \& Levinson, A., 2003, ApJ, 584, 937
\bibitem[van Putten \& Ostriker(2001)]{van01a} van Putten, M.H.P.M., \& Ostriker, E.C., 2001, ApJ, 552, L31 
\bibitem[van Putten(2008)]{van08a} van Putten, M.H.P.M., 2008, ApJ, 684, L91
\bibitem[Frontera et al.(2009)]{fro09} Frontera, F., Guidorzi, C., Montanari, E., et al., 2009, ApJ Suppl., 180, 192
\bibitem[Borgonov(2004)]{bor04} Borgonov, L., 2004, A\&A, 418, 487
\bibitem[Beloborodov et al. (1998)]{bel98} Beloborodov, A.M., Stern, B.E., \& Svensson, R., 1998, ApJ, 508, L25; ibid., 2000, ApJ, 535, 158
\bibitem[Jakobsson et al.(2005)]{jak05} Jakobsson, P., Levan, A., Fynbo, J.P.U.F., et al., 2005, A\&A, 447, 897
\bibitem[Guidorzi et al.(2012)]{gui12} Guidorzi, C., Margutti, R., Amati, L., et al., 2012, MNRAS, 422, 1785 
\bibitem[Caito et al.(2009)]{cai09} Caito, L., Bernardini, M.G., Bianco, C.L., Dainotti, M.G., Guida, R., and Ruffini, R., 2009, A\&A, 498, 501
\bibitem[Cuifolini \& Lavlis(2004)]{ciu04} Ciufolini, I., \& Pavlis, E.C., 2004, Nature, 431, 958; Everitt, C.W.F., et al., 2011, Phys. Rev. Lett., 106, 221101.
\bibitem[Everitt et al.(2011)]{eve11} Everitt, C.W.F., et al., 2011, Phys. Rev. Lett., 106, 221101
\bibitem[Kerr(1963)]{ker63} Kerr, R.P., 1963, Phys. Rev. Lett., 11, 237
\bibitem[Tavani(1996)]{tav96} Tavani, M., 1996, ApJ, 466, 768; M\'esz\'aros , P., \& Rees, M.J., 2000, ApJ, 530, 292; Liang, E., et al., 1997, ApJ, 479, L35; Blinnikov, S.I., Kozyreva, A.V., \& Panchenko, I.E., 1999, Astron. Rep., 43, 739; Lazatti, D., Ghisellini, G., Celotti, A., \& Rees, M.J., 2000, ApJ, 529, L17; Lazatti, D., Morsony, B.J., \& Begelman, M.C., 2009, ApJ, 700, L47
\bibitem[van Putten \& Levinson(2002)]{van02} van Putten, M.H.P.M., \& Levinson, A., 2002, Science, 295, 2874
\end{thebibliography}
\end{document}